\documentstyle[12pt,bezier]{article}
\makeatletter
\@addtoreset{equation}{section}
\makeatother

\newcommand{\VEV}[1]{\langle #1 \rangle}

\newcommand{\fsl}[1]{\not \kern-2pt #1} 
\newcommand{\DISP}{\displaystyle}

\newcommand{\integ}[2]{\int_{#1}^{#2}\!\!}

\newcommand{\GeV}{\hbox{GeV}}

\newcommand{\xbar}[1]{#1 \hspace{-5.5pt}/}

\newcommand{\SxSB}{\mbox{$S\chi SB$}}

\newcommand{\dfrac}[2]{\frac{\displaystyle{#1}}{\displaystyle{#2}}}

\def\fsl#1{\setbox0=\hbox{$#1$}           % set a box for #1 
   \dimen0=\wd0                                 % and get its size
   \setbox1=\hbox{/} \dimen1=\wd1               % get size of /
   \ifdim\dimen0>\dimen1                        % #1 is bigger
      \rlap{\hbox to \dimen0{\hfil/\hfil}}      % so center / in box
      #1                                        % and print #1
   \else                                        % / is bigger
      \rlap{\hbox to \dimen1{\hfil$#1$\hfil}}   % so center #1
      /                                         % and print /
   \fi}                                         %

\newcommand{\critline}{
\begin{picture}(150,150)(-15,-15)
  \thinlines
  \put(-10,0){\line(1,0){150}}
  \put(0,-10){\line(0,1){150}}
  \thicklines
  \put(100,-10){\line(0,1){35}}
  \put(107,-15){$\lambda_c$}
  \put(-2,100){\line(1,0){5}}
  \put(-15,107){1}
  \put(17,97){$g=g^{*}(\lambda)$}
  \bezier{300}(0,100)(99.7,50)(99.7,25)
  \put(25,25){$Sym.$}
  \put(75,75){\SxSB}
\end{picture}}
\begin{document}

\begin{titlepage}

\begin{flushright}
DPNU-96-09\\
January 1996\\
hep-ph/9603277
\end{flushright}
\begin{center}

\renewcommand{\thefootnote}{\fnsymbol{footnote}}
{\Large \bf
Top Quark Condensate Revisited%
\footnote{To appear in Proc. YKIS'95, 
  ``From the Standard model to Grand Unified Theories'', YITP, 
  Kyoto University, Kyoto, August 21-25, 1995
  (Supplement of Prog. Theor. Phys., 1996), ed. T. Kugo.}
 }
%\thanks{
 %   Work supported in part by the Sumitomo Foundation and a Grant-in Aid for
 %   Scientific Research from the Ministry of Education, Science and Culture 
 %   (No. 05640339).
 % }\\par

\renewcommand{\thefootnote}{\fnsymbol{footmote}}
\setcounter{footnote}{0}
\par
\vspace{1cm}
%\author{
{\large 
 {\bf  Koichi Yamawaki }\par
}
  Department of Physics, Nagoya University \\
  Nagoya 464-01, Japan

\vspace{1cm}

%\date{}

%\maketitle
\end{center}
\begin{abstract}
  %\small
  %\setlength{\baselineskip}{12pt}
  In the light of recent discovery of a very heavy top quark, we 
  reexamine the top quark condensate model 
  proposed by Miransky, Tanabashi and Yamawaki (MTY)
   and by Nambu. We first review the original MTY 
   formulation based on the ladder Schwinger-Dyson
     equation and the Pagels-Stokar formula.  
It is particularly emphasized that 
the critical phenomenon gives a simple reason why the
top quark can have an extremely large mass compared with other quarks
and leptons.
 Then we discuss the Bardeen-Hill-Lindner (BHL) formulation 
 based on the renormalization-group equation and 
  the compositeness condition, which 
  successfully picks up $1/N_c$-sub-leading effects disregarded by MTY.
  In fact BHL is equivalent to MTY at the $1/N_c$-leading order.
 Such a simplest version of the model predicts the top quark mass, $m_t
 \simeq 250 {\rm GeV}$ (MTY) and $m_t \simeq 220 {\rm GeV}$ (BHL),
 for the cutoff on the Planck scale. 
 In this version we cannot take the cutoff beyond
 the Landau pole of $U(1)_Y$ gauge coupling, 
 which yields a minimum
 value of the top mass prediction $m_t \simeq 200 {\rm GeV}$.
 We then propose a ``top mode walking GUT'' : The standard 
 gauge groups are unified into a (``walking'') GUT so that the cutoff
 can be taken to infinity 
 thanks to the renormalizability of the four-fermion theory coupled to
  ``walking'' gauge theory. 
 The top and Higgs mass prediction is then  
 controlled by the Pendleton-Ross infrared
 fixed point at GUT scale and can naturally 
 lead to $m_t \simeq m_H \simeq 180  {\rm GeV}$.
 \par
\end{abstract}

\end{titlepage}

\section{Introduction}

As it stands now, the standard model (SM) is a very successful framework for 
describing elementary particles in the low energy region, say, less than 
100 GeV. However one of the most mysterious parts of the theory, 
{\it the origin of mass}, has long been left unexplained. Actually, 
mass of {\it all} particles in the SM is attributed 
to a {\it single} order parameter, the vacuum expectation value of 
the Higgs doublet ($\simeq$ 250 GeV). Thus the problem of the origin of
mass is simply reduced to understanding dynamics of the Higgs sector.

Recently the elusive top quark has been finally discovered and found to have
a mass of about 180 GeV,\cite{kn:TEVATRON} roughly on the order of weak scale 
250 GeV. This is extremely large compared with mass of
all other quarks and leptons and seems to suggest a special role of the top
quark in the electroweak symmetry breaking, {\it the origin of mass},
and hence a strong connection with the Higgs boson itself.  
 
Such a situation can be most naturally understood by the top quark
condensate 
proposed by Miransky, Tanabashi and Yamawaki (MTY)\cite{kn:MTY89a,kn:MTY89b}
 and
by Nambu\cite{kn:Namb89} independently. This entirely replaces
the standard Higgs doublet by a composite one formed by a strongly
coupled short range dynamics (four-fermion interaction) which triggers
the top quark condensate.  The Higgs boson emerges as
a $\bar t t$ bound state and hence is deeply connected with the top
quark itself. Thus the model may be called ``top mode standard
 model''\cite{kn:MTY89b}
  in contrast to the SM (may be called ``Higgs mode standard
 model''). The model was further developed by the renormalization-group 
 (RG) method.\cite{kn:Marc89,kn:BHL90}
 
 Once we understand that the top quark mass is of the weak scale order,
 then the question is why other quarks
 and leptons have very small mass compared with the weak scale. Actually,
 the Yukawa coupling is dimensionless and hence naturally 
 expected to be of $O(1)$. This is the question that MTY 
 \cite{kn:MTY89a,kn:MTY89b} solved 
  in the top quark condensate through the amplification of the
  symmetry violation in the critical phenomenon.

MTY\cite{kn:MTY89a} introduced explicit 
four-fermion interactions responsible
for the top quark condensate in addition to the standard 
gauge couplings. 
Based on the explicit solution of the ladder
SD equation\cite{kn:KMY89,kn:MY89}
and the Pagels-Stokar (PS) formula\cite{kn:PS79},
MTY predicted
 the top quark mass to be about 250 GeV (for the Planck
scale cutoff), which actually coincides with the weak scale.
MTY also found that even if all the dimensionless four-fermion
couplings are of $O(1)$, only the coupling larger than the critical 
coupling yields non-zero (large) mass, while others do just zero masses.
  This is a salient feature of the {\it critical phenomenon}.
It should be emphasized that the MTY prediction 
(receipt date: Jan. 3, 1989)\cite{kn:MTY89a} 
was made when the lower bound of the top quark mass through direct experiment
was only 28 GeV (TRISTAN value) and many theorists (including SUSY enthusiasts) were still expecting the value below 100 GeV. 
It in fact appeared absurd at that time
to claim a top mass on the order of weak scale. 
Thus such a large top mass was really a {\it prediction} of the model.

The model was further formulated in an elegant
fashion by Bardeen, Hill and Lindner (BHL)\cite{kn:BHL90} in the
SM language, based on the RG equation 
and the compositeness condition. BHL incorporated composite
Higgs loop effects as well as the $SU(2)_L \times U(1)_Y$ gauge boson loops. 
Such effects turned out to reduce the above MTY 
value 250 GeV
down to 220 GeV, a somewhat smaller value but still on the
order of the weak scale.  Although the prediction appears to be
substantially higher than the experimental value  mentioned
above, there still remains a
possibility that (at least) an essential feature of the top quark
condensate idea may eventually survive.

In this talk we reexamine the simplest version of the top quark condensate 
in view of the recent discovery of a heavy top quark \cite{kn:TEVATRON}. 
We first review the top quark condensate model based 
on the explicit four-fermion interactions introduced 
by MTY\cite{kn:MTY89a,kn:MTY89b}. Combined with the standard gauge
interactions, dynamics of the model becomes a gauged Nambu-Jona-Lasinio (NJL) 
model. We then explain the MTY analysis of 
the model done in the ladder SD equation at the $1/N_c$ leading order. 
We shall emphasize how a {\it critical phenomenon} implied by the 
solution naturally explains why quarks and leptons other than the top quark
can have extraordinarily small mass compared with the weak scale.

 As to concrete mass
prediction, solution of the SD equation should be combined with 
the PS formula\cite{kn:PS79} for the decay constant of the composite 
Nambu-Goldstone (NG) bosons, $F_{\pi}\simeq 250 {\rm GeV}$, 
which determines the overall 
scale of the solution, namely the top mass.
We then explain the BHL\cite{kn:BHL90} formulation based on the
RG equation and the compositeness condition. It
essentially incorporates $1/N_c$ sub-leading effects such as
those of the composite Higgs loops and $SU(2)_L\times U(1)_Y$ gauge boson 
loops which were disregarded by the MTY formulation. We shall
 explicitly see that BHL is in fact  
equivalent to MTY at $1/N_c$-leading order. 
As far as the cutoff is below the Planck scale, 
the top quark mass prediction has a lower bound: $m_t
 \simeq 250 {\rm GeV}$ (MTY) or $m_t \simeq 220 {\rm GeV}$ (BHL).

We shall next experiment with the idea 
of taking the cutoff beyond the Planck scale.
 In this simplest version,
 even if we were allowed to ignore the quantum gravity effects, 
  we cannot take the cutoff beyond
 the Landau pole of $U(1)_Y$ gauge coupling, 
 which actually yields an absolute minimum
 value of the top mass prediction $m_t \simeq 200 {\rm GeV}$.
 However, if the standard 
 gauge groups are unified into a (``walking'')\footnote{
 Nowadays, a walking\cite{kn:YBM86}
  coupling means a {\it very} slowly running coupling with 
 $A=c/b \gg 1$, where $b,c$ are coefficients of the one-loop beta function
 and the anomalous dimension, respectively; $\beta(g)=-b g^3$, $\gamma(g)
 =cg^2$. In the context of renormalizability
 of the gauged NJL model, we here use ``walking'' for $A>1$ 
 (slow running) instead of $A\gg 1$ (very slow running, 
 or ``standing'').\cite{kn:BMSY87}
 }  
 GUT,  we may take the cutoff to infinity 
 thanks to the renormalizability 
 arguments\cite{kn:KSY91,kn:KTY91,kn:Yama91,kn:Kras93,kn:KSTY94,kn:HKKN94}
  of the gauged NJL model with ``walking'' gauge coupling. 
  We shall consider this possibility 
  (``top mode walking GUT'')\cite{kn:ITY96} in which the top and Higgs mass 
 prediction is 
 controlled by the Pendleton-Ross (PG) infrared
 fixed point \cite{kn:PR81} at GUT scale and can naturally 
 lead to $m_t \simeq m_H \simeq 180  {\rm GeV}$.
\par

\section{Top Mode Standard Model}

\subsection{The Model}

 Let us first explain the original version of the top quark condensate model
  (top mode standard model) proposed by MTY\cite{kn:MTY89a,kn:MTY89b}
  based on explicit four-fermion interactions. 
The model
consists of the standard three families of quarks and leptons 
with the standard $SU(3)_C \times SU(2)_L \times U(1)_Y$ gauge 
interactions 
but {\it without Higgs doublet}. Instead of the standard Higgs sector 
MTY introduced $SU(3)_C \times SU(2)_L \times U(1)_Y$-invariant 
{\it four-fermion interactions among quarks and leptons}, the origin of 
which is expected to be a new physics not specified at this moment. 
The new physics specifies the ultraviolet (UV) scale (cutoff $\Lambda$) of
 the model, in contrast to the infrared (IR) scale (weak scale $F_{\pi}\simeq 
 250 \GeV$)
 determined by the mass of $W/Z$ bosons.

The explicit form of such four-fermion 
interactions reads: \cite{kn:MTY89a,kn:MTY89b}   
\begin{eqnarray}
  {\cal L}_{4f}
  &=&
   \bigg[
       G^{(1)} 
         (\bar\psi_L^i \psi_R^j)
         (\bar\psi_R^j \psi_L^i)
  \nonumber \\
  & & +G^{(2)} 
         (\bar\psi_L^i \psi_R^j)
         (i\tau_2)^{ik}(i\tau_2)^{jl}
         (\bar\psi_L^k \psi_R^l)
  \nonumber \\
  & &+ G^{(3)} 
         (\bar\psi_L^i \psi_R^j)
         (\tau_3)^{jk}
         (\bar\psi_R^k \psi_L^i)
   \bigg]
     +\hbox{h.c.},
\label{eq:(4.1)}
\end{eqnarray}
where $i,j,k,l$ are the weak isospin indices
 and $G^{(1)}$, $G^{(2)}$ and 
$G^{(3)}$ are the four-fermion coupling constants
among top and bottom quarks $\psi\equiv (t,b)$.
It is straightforward\cite{kn:MTY89a,kn:MTY89b} to include other families and
leptons into this form.

The symmetry structure (besides $SU(3)_C$) of the 
four-fermion interactions, $G^{(1)}$, $G^{(2)}$ and 
$G^{(3)}$, is 
$SU(2)_L \times SU(2)_R \times U(1)_V \times U(1)_A$, 
$SU(2)_L \times SU(2)_R \times U(1)_V $ and 
$SU(2)_L \times U(1)_Y \times U(1)_V \times U(1)_A$, respectively. 
The $G^{(2)}$ term is vital to the mass of
the bottom quark in this model.\cite{kn:MTY89a,kn:MTY89b}
In the absence of the $G^{(2)}$-term, (\ref{eq:(4.1)}) possesses a 
$U(1)_A$ symmetry 
which is explicitly broken only by the color anomaly and plays the role
of the Peccei-Quinn symmetry.\cite{kn:MTY89b}

Let us disregard the $G^{(2)}$ term for the moment, in which case
the MTY Lagrangian (\ref{eq:(4.1)}) simply reads
\begin{equation}
  {\cal L}_{4f}
   =G_t (\bar\psi_L t_R)^2 + G_b (\bar\psi_L b_R)^2 + h.c.,
\label{eq:tb-4fermi}
\end{equation}
with $G_t \equiv G^{(1)}+G^{(3)}$ and $G_b \equiv G^{(1)}-G^{(3)}$.
The above MTY Lagrangian with $G_b=0$ was the starting point of
BHL \cite{kn:BHL90} but setting $G_b=0$ overlooks an important
aspect of the top quark condensate, as we will see in the followings. 

\subsection{Why $m_t \gg m_{b,c,\cdot\cdot\cdot}$?}

We now explain one of the key points of the model, i.e.,
explicit dynamics  which gives rise to a large isospin
violation in the condensate $\VEV{\bar tt}\gg \VEV{\bar bb}$
 ($m_t \gg m_b$), or more generally, naturally explains why
 only the top quark has a very large mass.
  MTY found\cite{kn:MTY89a,kn:MTY89b} that critical phenomenon, or
  theory having nontrivial UV fixed point with
  large anomalous dimension, is actually such a dynamics, based on the 
spontaneous-chiral-symmetry-breaking (S$\chi$SB) solution
of the ladder SD equation for the gauged NJL model.
For the $SU(3)_c\times SU(2)_L\times U(1)_Y$-gauged NJL model, 
the ladder SD equation becomes simpler in the large $N_c$ limit: 
Rainbow diagrams of
the $SU(2)_L\times U(1)_Y$ gauge boson lines are suppressed compared with
those of the QCD gluon lines. 

 For simplicity we first consider 
 the ladder SD equation with the {\it non-running} QCD coupling 
 and four-fermion coupling (\ref{eq:tb-4fermi}).
Without $G^{(2)}$ term, the
top and bottom quarks satisfy decoupled SD equations.
(We can easily find a solution for the SD equation with the $G^{(2)}$ 
term.)\cite{kn:MTY89b}
In Euclidean space, the ladder SD equation for each quark  
propagator $S_i^{-1}(p)=A_i(p^2)(\xbar{p} -\Sigma_i(p^2)) (i=t,b)$
in Landau gauge takes the form (after angular integration):
\begin{equation}
  \Sigma_i(x)
  = {g_i \over \Lambda^2} \int_0^{\Lambda^2} dy 
      {y\Sigma_i(y) \over y+\Sigma_i(y)^2}
    +\int_0^{\Lambda^2} dy
      {y\Sigma_i(y) \over y+\Sigma_i(y)^2} K(x,y),
\label{eq:(4.2)}
\end{equation}
where $x \equiv p^2 $, $K(x,y) =\lambda/\max(x,y)$ and $\Lambda$ is 
a UV cutoff (a scale of new physics). We have defined
 {\it dimensionless} four-fermion couplings
$
g_i \equiv (N_c\Lambda^2 /4\pi^2 )G_i 
$
and
$
\lambda \equiv (3C_2(\mbox{\boldmath$F$})/4\pi)\alpha_{QCD}
$, with 
$N_c(=3)$
and
$
C_2(\mbox{\boldmath$F$})=(N_c^2-1)/2N_c (=4/3)
$
being the number of color and 
the quadratic Casimir of the fermion color representation 
$\mbox{\boldmath$F$} (= {\bf 3)}$, respectively.
(Note that $A_i(p^2)=1$ in Landau gauge in the ladder approximation.)
The dynamical mass function is normalized as $\Sigma_i(m_i^2) = m_i$.

Before discussing reality, let us look at a simplified case with
the QCD coupling being switched off ($K(x,y)=0$). Then this SD equation is
 simply reduced to the {\it gap equation} of the 
NJL model\cite{kn:NJL61} in the large $N_c$ limit. 
It is well known that the NJL model has a nontrivial solution 
$
\Sigma(p^2)\equiv {\rm const.}= m \ne 0
$
for 
$
g>g^*=1.
$
In the critical region ($0< m/ \Lambda \ll 1$)
the solution reads (``scaling relation''):
\begin{equation}
\left({m \over \Lambda}\right)^2 \simeq \frac{1}{g^*} - \frac{1}{g}
\label{scalingNJL}
\end{equation}
up to logarithm.
In view of the existence of such a critical coupling $g=g^*=1$, it is easy
 to see that our four-fermion interactions (\ref{eq:tb-4fermi})
 can give a {\it maximal isospin violation} in dynamical mass;
  $m_t \ne 0$ and $m_b =0$,
 if $g_t> g^* =1 >g_b$ ({\it not necessarily} $g_t \gg g^* \gg g_b$). 
 Eq.(\ref{scalingNJL}) shows that 
  the dynamical mass sharply rises 
 to the order of cutoff, $m = O (\Lambda)$, as the coupling moves off
 the critical point.\footnote{
 This implies that we need a fine-tuning of bare coupling $1/g^* -1/g\ll 1$
 in order to guarantee a hierarchy $m\ll \Lambda$. In the pure NJL model
 the limit $\Lambda/m \rightarrow \infty$ leads to a trivial 
 (non-interacting) theory and hence this fine-tuning is not connected with
 a finite renormalized theory. In the gauged NJL model (with ``walking''
 gauge coupling, $A>1$), on the other hand,
 this fine-tuning is traded for a renormalization procedure arriving at
 a finite continuum (renormalized) theory defined at the UV fixed point
 with large anomalous dimension.\cite{kn:KSY91,kn:KTY91} 
 We shall later return to this 
 renormalizability
 of the gauged NJL 
 model.\cite{kn:KSY91,kn:KTY91,kn:Yama91,kn:Kras93,kn:KSTY94,kn:HKKN94}
 }  
 Thus the critical phenomenon distinguishes the
  top quark ($g_t >g^*$) from all others ($g<g^*$) {\it qualitatively}:
  $m_t \ne 0, m_{\rm others}=0$, even if all the couplings are 
  $O(1)$. 
 
Now, a similar argument applies to the gauged NJL model,
(\ref{eq:(4.2)}). The same type of equation as (\ref{eq:(4.2)})
 was first studied by Bardeen, Leung and
Love\cite{kn:BLL86} in QED for the strong gauge coupling region
$\lambda>\lambda_c =1/4$. A full set of spontaneous chiral
symmetry breaking (S$\chi$SB) solutions 
in the whole ($\lambda,g$) plane and the {\it critical line}
were found by Kondo, Mino and Yamawaki and
independently by Appelquist, Soldate, Takeuchi and
Wijewardhana.\cite{kn:KMY89} 

The critical line in the ($\lambda,g$) plane is a generalization of the
 critical coupling in NJL model. It is the line of the second-order phase
transition separating spontaneously broken ($m/\Lambda\ne0$)
and unbroken ($m/\Lambda=0$) phases of the chiral symmetry
(Fig.1)\cite{kn:KMY89}:
\par
\begin{figure}
\begin{center}
\critline \\
\begin{minipage}{5in}
  {\footnotesize
    {\bf Fig.1 \quad} 
  Critical line in $(\lambda,g)$ plane. It 
  separates spontaneously broken (\SxSB) phase and
  unbroken phase ($Sym.$) of the chiral symmetry.
}
\end{minipage}
\end{center}
\end{figure}
\begin{eqnarray}
  g
  &=&{1 \over 4}(1+\omega)^2
  \equiv g^{*},
  \qquad
  \omega \equiv \sqrt{1-\lambda/\lambda_c}
  \qquad (0< \lambda < \lambda_c), 
   \nonumber \\
  \lambda
  &=& \lambda_c  \qquad (g<{1\over4}). 
\label{eq:(2.17)}
\end{eqnarray}
The asymptotic form of the solution of the ladder SD 
equation (\ref{eq:(4.2)}) takes the form:\cite{kn:KMY89}
\begin{equation}
  \Sigma(p^2)\mathop{\simeq}_{p\gg m} 
%   {m^3\over p^2} \left({p\over m}\right)^{\gamma_m} 
    \simeq m\left({p\over m}\right)^{-1+\omega}
    \mathop{\simeq}_{\lambda\ll 1} m\left({p\over m}\right)^{-2\lambda},
\label{eq:(4.3)}
\end{equation} 
which is reduced to a constant mass function $\Sigma (p^2)\equiv 
{\rm const.}=m$ in the pure NJL limit ($\lambda \rightarrow 0$) as it
should be.
Through operator product expansion and RG equation,  
\begin{equation}
  \Sigma(p^2)\mathop{\simeq}_{p\gg m} 
   {m^3\over p^2} \left({p\over m}\right)^{\gamma_m},
\end{equation}
such a slowly damping solution (\ref{eq:(4.3)}) actually corresponds to 
 a large anomalous dimension: \cite{kn:MY89}
\begin{equation}
\gamma_m =1+\omega \qquad (0< \omega < 1),
\label{eq:gamma_m}
\end{equation}
at the critical line for $0< \lambda < \lambda_c$.

Here the overall mass scale $m$
at $0<\lambda<\lambda_c$ satisfies the form similar to 
(\ref{scalingNJL}):\cite{kn:KMY89}
\begin{equation}
\left({m \over \Lambda}\right)^{2\omega} \simeq \frac{1}{g^*}
 - \frac{1}{g}
\label{scalinggNJL}.
\end{equation}
As in the pure NJL model the dynamical mass $m$ sharply rises
as we move away from the critical coupling. 
Now the critical coupling of $g$ 
on the critical line does depend on
the value of gauge coupling $\lambda$ and vice versa. 
This means that {\it even 
a tiny difference (symmetry violation) of $\lambda$ ($g$) for the same $g$
($\lambda$) can cause amplified effects on the dynamical mass; $m=0$
(below the critical line) or $m\ne 0$ (above the critical line).}
 
Returning to the top quark condensate, we note that
our SD equations (\ref{eq:(4.2)}) separately include  
isospin-violating four-fermion 
couplings $g_t \not = g_b $ $(g^{(3)} \not = 0 ) $. 
In view of the critical line (\ref{eq:(2.17)}) and the critical
behavior (\ref{scalinggNJL}),
MTY\cite{kn:MTY89a,kn:MTY89b} indeed found  
{\it amplified
 isospin symmetry violation} for a small ({\it however small}) violation
 in the coupling constants. 
Thus we have an S$\chi$SB
solution with {\it maximal isospin violation}, $m_t \not =0$ and
 $m_b=0$,
when
\begin{equation}
 g_t> g^*={1\over 4}(1+\omega)^2 >g_b
\label{eq:tbsplit}
\end{equation}
($g_t$ is above the critical line and $g_b$ is below
it). As already mentioned, we need {\it not} to set $G_b =0$ 
in the four-fermion
 interactions (\ref{eq:tb-4fermi}) to obtain $m_b=0$. 
 Thus, even if we assume that all the dimensionless couplings are 
 $O(1)$,  the 
 {\it critical phenomenon}
 naturally explains why only the top quark can have a large mass,
 or more properly, {\it why other fermions can have very small masses}: 
 $m_t \gg m_{b,c,\cdot\cdot\cdot}$. It is indeed realized if
  only the top quark coupling is above the
 critical coupling, while all others below it: $g_t >g^*>
 g_{b,c,\cdot\cdot\cdot} \Longrightarrow m_t \ne 0, m_{b,c,\cdot\cdot\cdot}=0$.
Note that other couplings do {\it not} need to be zero nor {\it very} small.

\subsection{Running QCD Coupling}

One can easily take account of {\it running effects} of the QCD
 coupling in the ladder SD equation (``improved ladder SD 
equation'')\cite{kn:Higa84}
by replacing $\lambda$ in (\ref{eq:(4.2)}) by the one-loop running
 one $\lambda(p^2)$ parameterized as follows:
\begin{equation}
  \lambda(p^2) 
  =\cases{
    \lambda_{\mu} 
       & ($p^2 <\mu_{IR}^2$) \cr
       {A/2 \over \ln{( p^2/  \Lambda_{QCD}^2)}}   
       & ($p^2 >\mu_{IR}^2$) \cr
},
\label{eq:(3.19)}
\end{equation}
where $A =c/b=18C_2(\mbox{\boldmath$F$})/(11N_c-2N_f)$ ($=24/(33-2N_f)$)
 and 
$\lambda_{\mu} 
(= (A/2)/\ln{(\mu_{IR}^2/\Lambda_{QCD}^2)})
$ are constants and $\mu_{IR} (= O(\Lambda_{QCD})$ an
artificial ``IR cutoff'' of otherwise divergent running coupling constant
(We choose $\lambda_{\mu} > 1/4$ so as to trigger the S$\chi$SB already in
the pure QCD). 
Then the SD equation takes the form 
\begin{equation}
  \Sigma_i(x) = 
     {g_i \over \Lambda^2} 
      \integ0{\Lambda^2}dy{y \Sigma_i(y) \over y+\Sigma_i^2(y)}
    + \integ0{\Lambda^2} dy {y \Sigma_i(y) \over y+\Sigma_i^2(y)}
      {\mbox{\boldmath$K$}(x,y)},
\label{eq:(4.4)}
\end{equation}
where ${\mbox{\boldmath$K$}}(x,y) \equiv
 \lambda(\max(x,y,\mu_{IR}^2))/\max(x,y)$. Note that the
 non-running case is regarded as the ``standing'' limit
  $A \rightarrow \infty$ (with $\lambda_{\Lambda} \equiv \lambda(\Lambda^2)$
fixed) of the walking coupling ($A \gg 1$).\cite{kn:BMSY87}
 
The S$\chi$SB solution of (\ref{eq:(4.4)}) is logarithmically
 damping\cite{kn:MY89}, essentially the same as 
 (\ref{eq:(4.3)}) with the small power $\lambda (\sim \lambda_{\Lambda})
 \ll 1$:
\begin{equation}
  \Sigma(p^2) \simeq m \biggl[{\lambda(p^2) \over \lambda(m^2)}
                       \biggr]^{{A \over 2}},
                       \quad
                       A= {8\over7}  \quad (N_f=6).
\label{eq:(4.5)}
\end{equation}
In the case of pure QCD ($g=0$), such a very slowly damping solution
 (``irregular asymptotics'') is the {\it explicit} chiral-symmetry-breaking
 solution due to the quark bare mass.\cite{kn:Lane74,kn:Higa84}
  However, Miransky and
 Yamawaki\cite{kn:MY89}
 pointed out that it can be the S$\chi$SB solution {\it in the presence of an
 additional four-fermion interaction.} 
The solution corresponds to a very
 large anomalous dimension $\gamma_m \simeq 2-2\lambda_{\Lambda}$
 (compare with (\ref{eq:gamma_m})) near
 the ``critical line''\cite{kn:Take89,kn:KSY91}
\begin{equation}
 g=g^* \simeq 1-2\lambda_{\Lambda}
\label{eq:``criticalline''}
\end{equation}
at $\lambda_{\Lambda} \ll 1$.
(There is no
critical line in the rigorous sense in this case, since S$\chi$SB 
takes place in the whole coupling region due to pure QCD dynamics, yielding
dynamical mass $m=m_{QCD}= O(\Lambda_{QCD})$.)  
Note that 
(\ref{eq:``criticalline''}) coincides with the critical line in the
 non-running case (\ref{eq:(2.17)}), $g={1 \over 4}(1+\omega)^2
\simeq 1-2\lambda$, at $\lambda \ll 1$. Actually, we can obtain
exact expression for ``critical line'', which becomes identical
with the entire critical line (\ref{eq:(2.17)}) in the limit $A \rightarrow
 \infty$.\cite{kn:KSY91}

In view of the ``critical line'', we again have an S$\chi$SB solution with
{\it maximal isospin violation}, $m_t \ne 0 , m_b=0$ (apart from $m_{QCD}$)
, under a condition 
similar to (\ref{eq:tbsplit}); 
$
   g_t> g^* (\simeq 1-2\lambda_{\Lambda}) > g_b.
$

\section{Top Quark Mass Prediction}

\subsection{SD Equation plus PS formula (MTY)}

Now we come to the central part of the model, namely,
relating the dynamical mass of the condensed fermion
 (top quark) to the mass of $W/Z$ bosons.
 
The top quark condensate $\VEV{\bar tt}$ indeed yields a standard gauge
 symmetry breaking pattern
$SU(2)_L \times U(1)_Y \rightarrow U(1)_{em}$ 
to feed the mass of W and Z bosons.
Actually, the mass of $W$ and $Z$ bosons in the top quark condensate
 is generated via dynamical Higgs mechanism as 
in the technicolor:
\begin{equation}
  m_W^2=({g_2 \over 2}F_{\pi^{\pm}})^2,
  \qquad       m_Z^2 \cos^2\theta_W =({g_2 \over 2}F_{\pi^0})^2,
\label{eq:W/Zmass}
\end{equation}
  where $g_2$ is the $SU(2)_L$ gauge coupling, and $F_{\pi^{\pm}}$
    and $F_{\pi^0}$ are the decay constants
    of the composite NG bosons $\pi^{\pm},\pi^{0}$ to be absorbed
     into $W$ and $Z$ bosons, respectively. $F_{\pi} (\simeq 250{\rm GeV})$
determines the IR scale of the model and plays a central role in 
fixing the top quark mass.
    
Decay constants of those composite NG bosons may be calculated 
in terms of the Bethe-Salpeter (BS) amplitude of the NG bosons 
determined by the BS equation, which must be solved consistently with 
the SD equation for the fermion propagator.\cite{kn:ABKMN90}
Instead of solving the BS equation, however, 
here we use the famous PS
 formula\cite{kn:PS79} which expresses the decay constants
in terms of dynamical mass function $\Sigma (p^2)$ of the condensed 
fermion, i.e., a solution of the ladder SD equation (\ref{eq:(4.5)}).
The PS formula was generalized by MTY\cite{kn:MTY89a} to 
the $SU(2)$-asymmetric case 
$m_t \not = m_b$ and $m_{t,b} \not =0$:
\begin{eqnarray}
  F_{\pi^\pm}^2 
    &=& 
    %\integ0{\Lambda^2}dx I_{\pm} \big( \Sigma_t,\Sigma_b \big)      = 
      {N_c\over 8\pi^2} \int_0^{\Lambda^2}dxx\cdot
    \nonumber \\
    & & \cdot
  {\DISP (\Sigma_t^2+\Sigma_b^2)-{x\over4}(\Sigma_t^2+\Sigma_b^2)^\prime
   +{x\over2}(\Sigma_t^2-\Sigma_b^2)
   \left[
      {\DISP 1+(\Sigma_t^2)^\prime \over \DISP x+\Sigma_t^2}
     -{\DISP 1+(\Sigma_b^2)^\prime \over \DISP x+\Sigma_b^2}
   \right]
   \over
   \DISP (x+\Sigma_t^2)(x+\Sigma_b^2)
  },
  \label{eq:(4.6)}\\
  F_{\pi^0}^2 
    %&=& 
    %\integ0{\Lambda^2} dx I_0 \big(\Sigma_t,\Sigma_b \big)
    %\nonumber \\
    &=& {N_c\over 8\pi^2} \int_0^{\Lambda^2}dxx\cdot
    \left[
      {\DISP \Sigma_t^2-{x\over 4}(\Sigma_t^2)^\prime
       \over
       \DISP (x+\Sigma_t^2)^2
      }
     +{\DISP \Sigma_b^2-{x\over 4}(\Sigma_b^2)^\prime
       \over
       \DISP (x+\Sigma_b^2)^2
      }
    \right].
 \label{eq:(4.7)}
\end{eqnarray}

Let us consider the extreme case, the maximal isospin
 violation mentioned above, $\Sigma_t(p^2) \ne 0$ and $\Sigma_b(p^2)=0$.
We further take a ``toy'' case switching off the gauge interactions: 
$\Sigma_t (p^2)\equiv {\rm const.}$ (pure NJL limit).
Then (\ref{eq:(4.6)}) and
(\ref{eq:(4.7)}) are both {\it logarithmically divergent} at
 $\Lambda/ m_t \rightarrow \infty$ 
{\it with the same coefficient}:
\begin{eqnarray}
  F_{\pi^\pm}^2 
  &=& {N_c \over 8\pi^2}m_t^2
  \left[
    \ln{\Lambda^2 \over m_t^2} +{1 \over 2}\right],
  \label{eq:F-NJL1}\\
  F_{\pi^0}^2 
  &=& {N_c \over 8\pi^2}m_t^2 \ln{\Lambda^2 \over m_t^2}.
\label{eq:F-NJL2}
\end{eqnarray}
Now, we could predict $m_t$ by fixing $F_{\pi^\pm}
\simeq 250{\rm GeV}$ 
so as to have a correct $m_W$ through
(\ref{eq:W/Zmass}). Actually,  (\ref{eq:F-NJL1}) determines $m_t$ as a
 {\it decreasing function of cutoff $\Lambda$}.
  The largest physically sensible $\Lambda$ (new physics scale) would be
 the Planck scale $\Lambda\simeq 10^{19}\GeV$ at which we have a minimum value
  $m_t \simeq 145\GeV$. If we take the limit 
 $\Lambda \rightarrow \infty$, we would have $m_t \rightarrow 0$,
 which is nothing but triviality
 (no interaction) of the pure NJL model: $y_t \equiv\sqrt{2}
  m_t/F_\pi \rightarrow 0$
 at $\Lambda \rightarrow \infty$. 

One might naively expect a disastrous weak isospin violation for the maximal 
isospin-violating dynamical mass, $m_t \ne 0$ and $m_b =0$..
 However, for 
 $\Lambda\gg m_t$, (\ref{eq:F-NJL1}) and (\ref{eq:F-NJL2}) yield 
$
F_{\pi^\pm} \simeq F_{\pi^0}
$
and
\begin{equation}
  \delta\rho 
  \equiv {F_{\pi^\pm}^2- F_{\pi^0}^2 \over  F_{\pi^{\pm}}^2}
   ={N_c m_t^2 \over 16\pi^2 F_{\pi^{\pm}}^2} 
   \simeq {1 \over 2 \ln{{\Lambda^2 \over m_t^2}}}
   \ll 1.
\label{eq:(4.8)}
\end{equation}
Then the problem of weak isospin 
 relation can in principle be solved {\it without custodial symmetry}.
Actually, the isospin violation 
$
F_{\pi^\pm} \ne F_{\pi^0}
$
in (\ref{eq:(4.6)}) and (\ref{eq:(4.7)})
solely comes from the different propagators having different 
$\Sigma_i(p^2)$, essentially the IR quantity, which becomes less important 
for $\Lambda \gg m$, since the integral is UV dominant.
This is the essence of the ``dynamical mechanism'' of MTY to
save the isospin relation $\rho \simeq 1$ without custodial 
symmetry.\footnote{
In the alternative formulation made by BHL\cite{kn:BHL90} 
 this dynamical consequence is tacitly
  incorporated into their assumption to take the {\it renormalizable form} for
  the effective theory of the composite Higgs (pure Higgs sector). Actually,
  it is impossible to write down a renormalizable pure Higgs Lagrangian having
  isospin violation $F_{\pi^\pm}\ne F_{\pi^0}$ (it is possible in the nonlinear
  sigma model). 
}

Now in the gauged NJL model, QCD plus four-fermion interaction
 (\ref{eq:tb-4fermi}), essentially the same mechanism  
 as the above is operative. Based on the very slowly damping 
 solution of the ladder SD equation (\ref{eq:(4.5)}) and 
 the PS formulas, (\ref{eq:(4.6)}) and
 (\ref{eq:(4.7)}), MTY\cite{kn:MTY89a,kn:MTY89b} predicted
$m_t$ and $\delta \rho$ as the {\it decreasing function of
 cutoff $\Lambda$}.
For the Planck scale cutoff $\Lambda\simeq 10^{19} \GeV$,
we have:\cite{kn:MTY89a,kn:MTY89b}\footnote{
One may substitute into (\ref{eq:(4.5)}) the {\it numerical} solution
( instead of the analytical one
(\ref{eq:(4.5)})) of the ladder SD equation 
(\ref{eq:(4.4)}), the result being the same as 
(\ref{eq:(4.15)}).\cite{kn:King90}
}\addtocounter{footnote}{1} 
\begin{equation}
        m_t \simeq 250 \GeV,
\label{eq:(4.15)}
\end{equation}
\begin{equation}
        \delta \rho \simeq 0.02 \ll 1.          
\label{eq:(4.16)}
\end{equation}
This is compared with the pure NJL case $m_t \simeq 145\GeV$: The QCD
corrections are quantitatively rather significant
 (As we will see
later, presence of the gauge
coupling will also change the qualitative feature of the theory
from a nonrenormalizable/trivial theory into a renormalizable/nontrivial
one.)\cite{kn:KSY91,kn:KTY91,kn:Yama91,kn:Kras93,kn:KSTY94,kn:HKKN94}

It will be more convenient to write an {\it analytical}
 expression for $F_{\pi}$.
  Neglecting the derivative terms with $\Sigma_t (x)'$ and using
  (\ref{eq:(4.5)}), we may approximate (\ref{eq:(4.6)}) as 
\begin{eqnarray}
F_{\pi}^2
  &\simeq&  {N_c\over 8\pi^2}\int_{m_t^2}^{\Lambda^2}dx{\Sigma_t^2 \over x}
  \nonumber \\
  &\simeq&  {N_c m_t^2 \over 16\pi^2}{A \over A-1}
               {(\lambda(m_t^2))^{A-1} - (\lambda(\Lambda^2))^{A-1}
                  \over   
                     (\lambda(m_t^2))^{A}}.
\label{eq:marciano}
\end{eqnarray}
This analytic expression was obtained by Marciano\cite{kn:Marc89} 
in the case of $A=8/7$ ($ N_f=6$), which actually reproduces the MTY 
prediction\cite{kn:MTY89a,kn:MTY89b}.

\subsection{RG Equation plus Compositeness Condition (BHL)}

Now, we explain the BHL formulation\cite{kn:BHL90}
 of the top quark condensate, which is based on the RG equation combined with
 the compositeness condition. 
BHL start with the SM Lagrangian
which includes explicit Higgs field at the Lagrangian level:
\begin{equation}
{\cal L}_{Higgs} = -y_t (\bar \psi_L^i t_R \phi _i + \mbox{h.c.})
 +  \left(D_\mu \phi ^\dagger\right)\left(D^\mu \phi \right)
  -   m_H^2 \phi ^\dagger\phi 
   -  \lambda_4 \left(\phi ^\dagger\phi \right)^2,
\label{eq:top51} 
\end{equation}
where $y_t$ and $\lambda_4$ are Yukawa coupling of the top quark and 
quartic interaction of the Higgs, respectively.
BHL imposed ``compositeness condition'' on $y_t$ and $\lambda_4$ in such a
way that (\ref{eq:top51}) becomes the {\it MTY Lagrangian} (\ref{eq:tb-4fermi})
(with $G_b=0$):
\begin{equation}
  \dfrac{1}{y_t^2}\rightarrow 0, \quad
   \dfrac{\lambda _4}{y_t^4}\rightarrow 0 \qquad 
    \mbox{as }    \mu \rightarrow \Lambda,
\label{eq:top53}
\end{equation}
where $\mu$ is the renormalization point above which the composite dynamics
are integrated out to yield an effective theory (\ref{eq:top51}).
Thus the compositeness condition implies divergence 
at $\mu = \Lambda $ of both the Yukawa
coupling of the top quark and the quartic interaction of the Higgs. 

Now, in the one-loop RG equation, the beta function of $y_t$ is given by
\begin{equation}
  \beta(y_t) = \dfrac{y_t^3}{(4\pi )^2}
              \left(N_c + \dfrac{3}{2}\right)
             - \dfrac{y_t}{(4\pi )^2}
                \left(3\frac{N_c^2-1}{N_c}g_3^2
                +\frac{9}{4}g_2^2-\frac{17}{12}g_1^2\right),
\label{eq:top61}
\end{equation}
where $g_1$, $g_2$ and $g_3$ are the gauge couplings of $U(1)_Y$, $SU(2)_L$
and $SU(3)_C$, respectively. BHL solved the RG equation for the
beta function (\ref{eq:top61}) combined with 
the compositeness condition (\ref{eq:top53}) as a boundary condition
at $\mu =\Lambda$.\\

\subsection{BHL versus MTY}

Let us 
first demonstrate \cite{kn:Yama91} 
that {\it in the large $N_c$ limit} BHL formulation \cite{kn:BHL90} 
is equivalent to that of MTY \cite{kn:MTY89a,kn:MTY89b}, both based on
the same MTY Lagrangian (\ref{eq:tb-4fermi}).
In the $N_c\rightarrow \infty$ limit for (\ref{eq:top61}),
we may neglect the factor $3/2$ in the first term 
(composite Higgs loop effects) and $g_2^2$ and $g_1^2$
in the second term (electroweak gauge boson loops), which
corresponds to the similar neglection of $1/N_c$ sub-leading
effects in the ladder SD equation in the MTY approach. 
Then (\ref{eq:top61}) becomes simply:
\begin{equation}
  \dfrac{d y_t}{d \mu}= \beta(y_t) = N_c \dfrac{y_t^3}{(4\pi )^2}  
                       - \dfrac{3N_c y_t g_3^2}{(4\pi )^2}.
\label{eq:top62}
\end{equation}
Within the same approximation the beta function of the 
QCD gauge coupling reads
\begin{equation}
  \dfrac{d g_3}{d \mu} = \beta (g_3) = -\frac{1}{A} 
  \dfrac{3N_c g_3^3}{(4\pi )^2}.
\label{eq:top63}
\end{equation}
Solving (\ref{eq:top62}) and (\ref{eq:top63}) by imposing 
the compositeness condition at $\mu =\Lambda $, we arrive at
\begin{equation}
  y_t^2(\mu ) = 
     \dfrac{2(4\pi )^2}{N_c} \frac{A-1}{A} 
     \dfrac{\left(\lambda (\mu^2 )\right)^A}
     {\left(\lambda (\mu^2 )\right)^{A-1}
     - \left(\lambda (\Lambda^2 )\right)^{A-1}}.
\end{equation}
Noting the usual relation $m_t^2=\frac{1}{2}y_t^2(\mu =m_t)v^2$
 $(v= F_{\pi})$,
we obtain 
\begin{equation}
 \dfrac{m_t^2}{F_{\pi}^2}
 %(\equiv g_Y^2)
   = \dfrac{y_t^2(m_t)}{2}=
     \dfrac{(4\pi )^2}{N_c} \frac{A-1}{A} 
     \dfrac{\left(\lambda (m_t^2 )\right)^A}
     {\left(\lambda (m_t^2 )\right)^{A-1}
     -\left(\lambda (\Lambda^2 )\right)^{A-1}}.
\label{BHLyukawa}
\end{equation}
This is precisely the same formula as (\ref{eq:marciano}) obtained in the
 MTY approach based on the SD 
 equation and the PS formula.\footnote{
 Alternatively, we may define $F_{\pi}^2(\mu^2) \equiv 2m_t^2/y_t^2(\mu)$
  which coincides 
 with the integral (\ref{eq:marciano}) with the IR end $m_t^2$ simply replaced 
  by $\mu^2$. Then the compositeness condition (\ref{eq:top53}) reads
  $F_{\pi}^2(\mu^2=\Lambda^2)=0$ (no kinetic term of the 
  Higgs).
  }
  Thus we have established
  \begin{equation}
  {\rm BHL} (\frac{1}{N_c} {\rm leading}) = {\rm MTY}.
  \end{equation}
 
Having established equivalence between MTY and BHL in the
large $N_c$ limit, we now
comment on the relation between them in more details.
 Note that MTY formulation is based on the nonperturbative picture,
 ladder SD equation and PS formula, which is valid at
 $1/N_c$ leading order, or the NJL bubble sum 
 with ladder-type QCD corrections
 (essentially the leading log summation).
 MTY extrapolated this $1/N_c$ leading picture all the
 way down to the low energy region where the sub-leading effects
 may become important. 
 
 On the other hand, BHL is crucially based on  
 the perturbative picture, one-loop RG equation, which can
 easily accommodate $1/N_c$ sub-leading effects 
 in (\ref{eq:top61}) such as 
 the loop effects of composite Higgs and electroweak gauge bosons.
 However, BHL formalism must necessarily be combined with the 
 compositeness condition (\ref{eq:top53}). The compositeness condition
  is obviously inconsistent with the perturbation and
 is a purely nonperturbative concept based on the same $1/N_c$ leading 
 NJL bubble sum as in the MTY formalism.
 Thus the BHL perturbative picture breaks down at high energy 
 near the compositeness scale $\Lambda$ where
 the couplings $y_t$ and $\lambda_4$ blow up as required by the 
 compositeness condition.

 So there must be a certain
 matching scale $\Lambda_{\rm Matching}$  such that 
 the perturbative picture (BHL) is valid for $\mu<\Lambda_{\rm Matching}$, 
 while 
 only the nonperturbative picture (MTY) 
 becomes consistent for $\mu >\Lambda_{\rm Matching}$.
 Such  a point may be defined by the energy region where the two-loop
 contributions dominate over the one-loop ones. A simple way to do such a
 matching is to use the BHL perturbative formalism for $\mu<
 \Lambda_{\rm Matching}$,
 while using the MTY formalism (or equivalently the BHL at $1/N_c$ leading 
 order) for $\mu >\Lambda_{\rm Matching}$.\footnote{
 Of course, the $1/N_c$ leading picture 
 might be subject to ambiguity such as the possible higher dimensional 
 operators, cutoff procedures, etc., all related with the nonrenormalizability
 of the NJL model.\cite{kn:Suzu90b} 
 These problems will be conceptually solved and 
 phenomenologically tamed, when coupled
 to the (``walking'' ($A>1$))
 gauge interactions (renormalizability of the
 gauged NJL 
 model)\cite{kn:KSY91,kn:KTY91,kn:Yama91,kn:Kras93,kn:KSTY94,kn:HKKN94} to be
 discussed later. Here we just comment that even if
 there might be such an ambiguity, the $1/N_c$ picture (MTY)
 is the only 
 consistent way to realize the compositeness condition as was done by
 the BHL paper itself.  
 }
 Thus the reality may in principle be expected to 
 lie in between BHL and MTY. 
 From $1/N_c$ sub-leading terms in (\ref{eq:top61}) 
 we can see that the composite Higgs
 loops push down the Yukawa coupling at low energy, while
 somewhat smaller effects of the electroweak gauge boson loops 
 make contributions in the opposite direction. As a result we would expect
 that BHL value is smaller than MTY one: 
 \begin{equation}
 m_t ({\rm BHL}) < m_t <m_t ({\rm MTY}).
 \end{equation}
 
 However, 
 thanks to the presence of a {\it quasi-infrared fixed point}\cite{kn:Hill81},
 BHL prediction is {\it numerically}
  quite stable against ambiguity at high energy region,
 namely, rather independent of whether this high energy region is 
 replaced by MTY or something else.
 Then we expect $m_t \simeq m_t ({\rm BHL}) =
 \frac{1}{\sqrt{2}} y_t(\mu=m_t) v \simeq \frac{1}{\sqrt{2}}\bar y_t v$ 
 within 1-2 \%, where 
$\bar y_t$ is the quasi-infrared fixed point given by $\beta(\bar y_t)=0$
in (\ref{eq:top61}). The composite Higgs loop changes $\bar y_t^2$ by
roughly the factor $N_c/(N_c+3/2)=2/3$ compared with the MTY value, i.e.,
$250 {\rm GeV}\rightarrow 250\times \sqrt{2/3}=204 {\rm GeV}$, 
while the electroweak gauge boson loop with opposite sign pulls it 
back a little bit to a higher value.
The BHL value\cite{kn:BHL90} is then given by
\begin{equation}
  m_t=218\pm3, \qquad \mbox{at } \Lambda \simeq 10^{19}\GeV.
\end{equation}

The Higgs boson was predicted as a $\bar t t$ bound state 
with a mass $M_H \simeq 2 m_t$\cite{kn:MTY89a,kn:MTY89b,kn:Namb89} 
based on the pure NJL model calculation.\cite{kn:NJL61}
Its mass was also calculated by BHL\cite{kn:BHL90} 
through the full RG equation of $\lambda_4$, the result being
\begin{equation}
  M_H = 239\pm3  (M_H/m_t\simeq 1.1)
   \qquad \mbox{at } \Lambda \simeq 10^{19}\GeV.
\end{equation}
If we take only the $1/N_c$ leading terms, we would have the mass ratio 
$M_H/m_t \simeq \sqrt{2}$, which is also obtained through the 
ladder SD equation.\cite{kn:STY90}

%\section{Cutoff Beyond Planck Scale?}

\section{Top Mode Walking GUT}

As we have seen, the top quark condensate naturally explains, through
the critical phenomenon, why only the top
quark mass is much larger than that of
other quarks and leptons: $m_t \gg m_{b,c,\cdot\cdot\cdot}$.
It further predicts the top mass on the order of weak scale. However, the 
predicted mass $220 {\rm GeV}$ is somewhat larger than the mass of the 
recently discovered top quark, $176 {\rm GeV}\pm 13 {\rm GeV}$ (CDF) and
$199 +38/-36 {\rm GeV}$ (D0)\cite{kn:TEVATRON}. 
Here we shall discuss a possible remedy of this
problem within the simplest model based on the MTY Lagrangian 
(\ref{eq:tb-4fermi}).\cite{kn:ITY96}

\subsection{Landau Pole Scenario}

First we recall that the top mass prediction is a {\it decreasing} function of 
the cutoff $\Lambda$. Then the simplest way to reduce the top mass would be
to raise the cutoff as much as possible. Let us assume
 that quantum gravity effects would not change drastically 
 the physics described by the low energy  theory without gravity.
 Then we may raise the cutoff $\Lambda$ beyond the Planck scale up to
 the Landau pole $\Lambda \simeq 10^{41} {\rm GeV}$ 
 where the $U(1)_Y$ gauge coupling $g_1$ diverges and the SM description 
 itself stops to be self-consistent.
In such a case the top and Higgs mass prediction becomes:
\begin{equation}
m_t \simeq 200 {\rm GeV}, \qquad  M_H \simeq 209 {\rm GeV}
\qquad \mbox{at } \Lambda\simeq 10^{41}\GeV
\end{equation}
which is the absolute minimum value of the prediction
within a simplest version of the top quark condensate.

If it is really the case, it would imply composite $U(1)_Y$ gauge
boson and composite Higgs generated {\it at once by the same dynamics}, 
since the Landau pole then may be regarded as a BHL compositeness condition 
also for the vector bound state as well as the composite Higgs.
Actually, we can formulate the BHL compositeness condition for
vector-type four-fermion interactions
 (Thirring-type four-fermion theory) 
as a {\it necessary condition} for the 
formation of a vector bound state. The possibility that both the
Higgs and $U(1)_Y$ gauge boson can be composites by the same dynamics
may be illustrated by an explicit model, the Thirring model in $D (2<D<4)$
dimensions. Reformulated as a gauge theory through hidden local 
symmetry, the Thirring model was shown
to have the dynamical mass 
generation, which implies that a composite Higgs 
and a composite gauge boson are generated at the same time.\cite{kn:IKSY95}

At any rate, the prediction of this scenario $m_t \simeq 200 {\rm GeV}$
still seems to be a little bit higher than the experimental
value, although the situation is not very conclusive yet.

\subsection{Renormalizability of Gauged NJL model}

Then we shall consider another possibility, namely, taking
 the cutoff to infinity: $\Lambda\rightarrow \infty$. 
In order to do this we should first discuss the renormalizability
of the gauged NJL model with ``walking'' gauge 
coupling 
($A>1$).\cite{kn:KSY91,kn:KTY91,kn:Yama91,kn:Kras93,kn:KSTY94,kn:HKKN94}

This phenomenon was first pointed out by Kondo, Shuto and 
Yamawaki\cite{kn:KSY91}
through the convergence of $F_{\pi}$ in the PS formula 
for the solution of the SD equation (\ref{eq:(4.5)}) in the
four-fermion theory plus QCD. 
 Contrary to the logarithmic divergence of (\ref{eq:F-NJL1})
in the pure NJL model, it was emphasized that 
for $A>1$ we have a {\it convergent} 
integral for $F_{\pi}$ and hence 
a nontrivial (interacting)
 theory $y_t\equiv m_t/F_{\pi}\ne 0$ in the continuum limit:
 Namely, the presence of ``walking'' ($A>1$) gauge interaction 
changes the trivial/nonrenormalizable theory (pure NJL model) into a 
nontrivial/renormalizable theory (gauged NJL model).\cite{kn:KSY91}

As to the non-running (standing) case ($A \rightarrow \infty$), 
the integral for 
$F_{\pi}^2$ is more rapidly convergent, since $\Sigma(p^2)$ is power
damping, (\ref{eq:(4.3)}), instead of logarithmic damping. 
In this case the renormalization
procedure was performed explicitly by Kondo, Tanabashi and 
Yamawaki \cite{kn:KTY91} through the effective 
potential in the ladder approximation.
The fine-tuning of the bare couling $1/g^*-1/g\ll 1$
in (\ref{scalinggNJL}) corresponds to the continuum limit 
$\Lambda/m \rightarrow \infty$, which now defines a finite
renormalized theory explicitly written in terms
of renormalized quantities, in sharp contrast to
the pure NJL model where the similar fine-tuning in (\ref{scalingNJL})
has nothing to do
with a finite renormalized theory. 
This renormalization led to the beta function and the anomalous dimension:
\begin{eqnarray}
  \beta(g) &=& 2 \omega g \left(1-\frac{g}{g^{*}} \right), 
\label{eq:beta}  \\
  \gamma_m(g) &=& 1-\omega + 2\omega \frac{g}{g^{*}},
\label{eq:gamma}  
\end{eqnarray}
where both functions take the same form in either bare or renormalized coupling $g$.
These expressions are valid both in the \SxSB and unbroken phases.
It is now clear that the critical line $g=g^*=\frac{1}{4}(1+\omega)^2$ is 
a UV fixed line where the anomalous dimension takes the 
large value (\ref{eq:gamma_m}):
\begin{equation}
1<\gamma_m(g=g^*) = 1+\omega<2.
\end{equation}
This result was first obtained by Miransky and Yamawaki \cite{kn:MY89}
for the bare coupling in the \SxSB phase, 
and was further shown \cite{kn:KTY91} to hold 
in both phases and also for the renormalized coupling, based on the effective 
potential.

The essence of the renormalizability now resides in the fact
that this dynamics possesses a large anomalous dimension
$\gamma_m>1$ but not too large, $\gamma_m <2$.\cite{kn:KTY91}
It in fact implies that 
 the four-fermion interactions are relevant operators,
  $2< d_{(\bar \psi \psi)^2}=2(3-\gamma_m)=4-2\omega <4$.\cite{kn:MY89}
Accordingly, possible {\it higher dimensional interactions},
  $(\bar \psi \psi)^4$, $\partial_\mu(\bar \psi  \psi)
  \partial^{\mu}(\bar \psi  \psi)$, etc.,
are {\it irrelevant operators} ($d >4$ due to $d_{\bar \psi \psi}>1$), 
 in contrast to the case without gauge interactions where these operators
are marginal ones ($d=4$ due to $d_{\bar \psi \psi}=1$)).

Returning to the ``walking'' coupling, we note that 
the anomalous dimension is given as $\gamma_m \simeq 2-2\lambda_{\Lambda}$
which is very close to 2 but less than 2 by only a logarithmic factor.
Then the above arguments for the standing coupling become rather delicate
in this case.
 In order to discriminate between $A>1$ and $A<1$, we again
  discuss the finiteness of $F_{\pi}$, or equivalently finiteness
  of effective Yukawa coupling, $y_t \equiv \sqrt{2}m_t/F_{\pi}>0$,
  in the continuum limit $\Lambda \rightarrow \infty$. 
 The analytical 
expression of the effective Yukawa coupling is already given by 
(\ref{eq:marciano}) (MTY), which is equivalent to  
 (\ref{BHLyukawa}) obtained as a solution of 
the RG equation with a compositeness condition at $1/N_c$ leading (BHL).
>From this expression it was noted\cite{kn:Yama91} 
that {\it iff $A>1$} (``walking'' gauge coupling with $N_c\sim N_f\gg 1$), 
then the effective Yukawa coupling remains finite, 
$y_t >0$, in the continuum limit $\Lambda \rightarrow \infty$.
This is in sharp contrast to the triviality of the
pure NJL model in which $y_t \rightarrow 0$ in the continuum
limit as was mentioned earlier. 

It was further pointed out
by Kondo, Tanabashi and Yamawaki\cite{kn:KTY91}
that this renormalizability is equivalent to existence of
a PR infrared fixed point\cite{kn:PR81} for the gauged
Yukawa model. The PR fixed point is given by the solution of
$\dfrac{d (y_t/g_3)}{d \mu}=0$ with 
(\ref{eq:top62}) and (\ref{eq:top63}):
\begin{equation}
y_t^2 
%=\frac{12\pi C_2({\bf F})}{N_c}\frac{A-1}{A}\alpha_{QCD}
      =\frac{(4\pi)^2}{N_c}\frac{A-1}{A}\lambda,
\label{PR}
\end{equation}
where $\lambda=3 C_2({\bf F}) g_3^2/(4\pi)^2$. Similar argument was recently 
developed more systematically by Harada, Kikukawa, Kugo and 
Nakano. \cite{kn:HKKN94}.

\subsection{Top Mode Walking GUT}

In view of the renormalizability
of the gauged NJL model with ``walking'' gauge coupling, we may take 
the $\Lambda \rightarrow \infty$ limit of the
top quark condensate.
However, in the realistic case we actually have the 
 $U(1)_Y$ gauge coupling which, as it stands, grows at high energy
 to blow up at Landau pole and hence 
 invalidates the above arguments of the renormalizability. Thus, in order to
apply the above arguments to the top quark condensate, 
we must remove the $U(1)_Y$ gauge interaction in such a way
as to unify it into a GUT with ``walking'' coupling ($A>1$)
beyond GUT scale. 
Then the renormalizability requires that the GUT coupling at GUT
scale should be determined by the PR infrared fixed point.\cite{kn:ITY96}

For a simple-minded GUT with $SU$-type group, the PR fixed point
takes the form similar to (\ref{PR}):
\begin{eqnarray}
y_t^2 (\Lambda_{\rm GUT}) 
&=& \frac{3 C_2({\bf F})}{N_c}\frac{A-1}{A} 
g_{\rm GUT}^2 (\Lambda_{\rm GUT})
          \simeq \frac{3}{2} g_{\rm GUT}^2 (\Lambda_{\rm GUT}), \\
\lambda_4 (\Lambda_{\rm GUT})
&=&\frac{6 C_2({\bf F})}{N_c}\frac{(A-1)^2}{A(2A-1)} 
g_{\rm GUT}^2 (\Lambda_{\rm GUT})
          \simeq \frac{3}{2} g_{\rm GUT}^2 (\Lambda_{\rm GUT}),
\end{eqnarray}
where we assumed $N_c\gg 1$ and $A\gg 1$ ($N_f\sim N_c\gg 1$) 
for simplicity. Then the 
top Yukawa coupling at GUT scale is essentially determined by the GUT 
coupling at GUT scale up to some numerical factor depending on the GUT
group and the representations of particle contents. Using 
``effective GUT coupling'' including such possible numerical factors, 
we may perform the BHL full RG equation analysis 
for $\mu <\Lambda_{\rm GUT} \simeq 10^{15} {\rm GeV}$
with the boundary condition of the above PR fixed
point at GUT scale.

For typical values of the effective GUT coupling $\alpha_{\rm GUT}\equiv 
g_{\rm GUT}^2/4\pi=1/40, 1/50, 1/60$,
prediction of the top and Higgs masses reads:
\begin{equation}
(m_t, M_H)\simeq (189, 193) {\rm GeV}, (183, 183){\rm GeV}, 
(177,173){\rm GeV},
\end{equation}
respectively. Note that these PR fixed point values at GUT scale 
are somewhat smaller than the coupling values at GUT scale 
which focus on the quasi-infrared fixed point in the low energy region.
Thus the prediction is a little bit away from the quasi-infrared fixed point. 
This would be the simplest extension of the top quark condensate 
consistent with the recent experiment on the top quark mass.

\section{Conclusion}

In the light of recent discovery of a very heavy top quark, 
we have reexamined the top quark condensate (top mode standard model).
A salient feature of the model is to give a simple explanation of
an extremely large mass of the top quark compared with other 
quarks and leptons {\it as a critical phenomenon} \cite{kn:MTY89a}. 
See Eq. (\ref{scalinggNJL}). Even if
dimensionless four-fermion couplings $g_t,g_b,\cdot\cdot\cdot$
are all $O(1)$, we still can have a large hierarchy $m_t\gg
m_{b,c,\cdot\cdot\cdot}$ {\it iff} there exists nonzero critical
coupling (critical line) $g^*$ such that
$g_t> g^* >g_{b,c,\cdot\cdot\cdot}$ (not necessarily
$g_t\gg g^* \gg g_{b,c,\cdot\cdot\cdot}$). This is an amplification
mechanism of symmetry violation characteristic to the critical phenomenon
(or, dynamical symmetry breaking having a nontrivial UV fixed point/line 
$g^*$ and a large anomalous dimension).

The original MTY \cite{kn:MTY89a} formulation 
predicted $m_t \simeq 250 {\rm GeV}$ (for Planck scale cutoff), based on a
purely nonperturbative picture of 
the large $N_c$ limit of the ladder SD equation and the PS formula,
i.e., the bubble sum with leading log 
QCD corrections. 
On the other hand, the BHL \cite{kn:BHL90} formulation 
reduced the above MTY value to $220 {\rm GeV}$,  
incorporating the $1/N_c$ sub-leading effects such as the composite 
Higgs loops and electroweak gauge boson loops, based on 
a combination of the (perturbative) RG equation and the (nonperturbative)
compositeness condition. In fact, if we pick up only
the $1/N_c$ leading order in BHL formulation, then it 
becomes equivalent to MTY. The perturbative picture of BHL 
breaks down in the high energy region near the compositeness scale where
the couplings blow up due to the very condition of the compositeness 
condition: The compositeness condition can only be justified nonperturbatively 
through the large $N_c$ argument. Thus the BHL formulation must in
principle be switched over to its  $1/N_c$ leading part, 
or equivalently the MTY formulation, in high energies.
As far as actual numerical prediction is concerned, however,
the above BHL value is quite insensitive to
this switch-over, thanks to the quasi-infrared fixed point.

Then we experimented with a heretic idea to raise the cutoff
scale beyond the Planck scale, ignoring all possible effects of
the quantum gravity which we do not know at present anyway.
First we simply placed the cutoff scale on 
the Landau pole of the $U(1)_Y$
gauge coupling, $\Lambda \simeq 10^{41} {\rm GeV}$,
the largest scale for which the top quark condensate with
the SM gauge couplings can be self-consistent. This yields
$m_t \simeq 200 {\rm GeV}$, which is 
absolutely the smallest value of the top mass 
within such a simplest version of the top quark condensate.

Next we considered a drastic possibility
 that the cutoff may be taken to infinity,
based on the renormalizability of the gauged NJL model.
In order to make the model renormalizable,
we should remove the $U(1)_Y$ factor by
unifying the SM gauge groups into a GUT with ``walking'' coupling
($A>1$). In this renormalizable ``top mode walking GUT''
in the infinite cutoff limit,
the couplings $y_t, \lambda_4$
at GUT scale are essentially given in 
terms of the GUT gauge coupling through
the PR infrared fixed point, which
can naturally predict somewhat
lower values of the top and Higgs masses:
$m_t \simeq M_H \simeq 180 {\rm GeV}$.  

Although the situation about top quark mass is still not yet conclusive,
we hope that at least essence of the idea of the top quark condensate
may eventually survive in the sense that the {\it origin of mass}
is deeply related to the top quark mass. 

\medskip

I would like to thank Iwana Inukai and Masaharu Tanabashi for
 collaborations and discussions on the recent results presented in this talk.
This work is supported in part
 by the Sumitomo Foundation and a Grant-in Aid for
    Scientific Research from the Ministry of Education, Science and Culture 
    (No. 05640339).

\end{document}